\documentclass[preprint,pra,aps,draft]{revtex4}

\usepackage{graphicx}
\usepackage{dcolumn}
\usepackage{bm}

\def\E{{\cal E}}
\def\x{{\bf x}}

\def\p{\mathbf{p}}

\def\a{\alpha}
\def\b{\beta}

\def\ih{{ \frac{i}{\hbar} }}
\def\au{{\underline \alpha}}
\def\bu{{\underline \beta}}

\def\E{{\cal E}}
\def\p{\partial}
\def\la{\langle}
\def\ra{\rangle}

\def\s{{\sigma}}
\def\a{\alpha}
\def\b{\beta}
\def\e{\epsilon}

\def\ih{{ {i \over \hbar} }}

\def\bx{{\bar x}}
\def\by{{\bar y}}
\def\p{\mathbf{p}}

\def\x{\mathbf{x}}

\def\R{{I\kern-0.3em R}}
\def\au{{\underline \alpha}}
\def\bu{{\underline \beta}}

\newcommand\be{\begin{equation}}
\newcommand\ee{\end{equation}}
\newcommand\bea{\begin{eqnarray}}
\newcommand\eea{\end{eqnarray}}

\begin{document}
\title{Some Recent Developments in the \\ Decoherent
Histories Approach to Quantum Theory}
\author{Jonathan Halliwell}
\affiliation{Blackett Laboratory, Imperial College, London, SW7
2BZ}

\vskip 1.0in

\begin{abstract}
A brief introduction to the decoherent histories approach to
quantum theory is given, with emphasis on its role in the
discussion of the emergence of classicality from quantum theory.
Some applications are discussed, including quantum-classical
couplings, the relationship of the histories approach to quantum
state diffusion, and the application of the histories approach to
situations involving time in a non-trivial way.
\end{abstract}

\maketitle

\section{Introduction}

Standard quantum theory is a remarkably successful theory. Indeed,
there is at present not one shred of experimental evidence that
its basic structure of Hilbert spaces, states, operators {\it
etc.} is in any way wrong. However, in its normal presentation,
the Copenhagen interpretation \cite{WZ}, it rests on certain
assumptions that may be restrictive. Firstly, it relies on a
division into classical and quantum domains. And secondly, it
places great emphasis on the notion of measurement \cite{Bell}.
These assumptions are not of course a restriction in the regimes
to which it is normally applied. Yet it invites the question, can
we do better? Can we formulate standard quantum theory in such a
way that it does not rely on these assumptions?

There are a number of reasons why one might want such a more general
formulation. Recent experiments have started to probe the
traditional border between what is normally called classical
and quantum \cite{Leg}. Furthermore, classical objects are after all built
out of atoms, which are quantum mechanical in many of
their properties. One can therefore ask how does classical behaviour
emerge on large scales for objects made out of small quantum
constituents? In addition, some current views of the early
universe, and in particular in the area of research
known as quantum cosmology, it is supposed that all force
and matter fields are subject to the laws of quantum
theory. There is no classical domain and there are certainly
no measuring devices. How then can we understand, in
a truly quantum-mechanical way, how measurements
made in the present epoch are related to events in the
distant past?

The decoherent histories approach to quantum theory is a
reformulation of standard quantum theory for closed quantum
systems (such as the entire universe), that removes the usual
emphasis on the notion of measurement and of a classical domain.
In this contribution I will briefly describe the formalism and the
way it explains the emergence of classical physics from an
underlying quantum theory. I will then also describe a selection
of applications of the decoherent histories approach in a variety
of different circumstances. This will show what sort of light the
approach has been able to shed on other fields. This review is by
no means an exhaustive survey of the field, which has by now
become quite extensive, and the choice of topics covered largely
reflects my own interests in the subject.

\section{Interference, Decoherence and Classicality}

Before embarking on a full discussion of the decoherent
histories formalism, let us first consider in a simple
way the phenomenon of interference in the double slit
experiment, its destruction through decoherence, and
the relationship of this to emergent classicality.

Consider then the standard double slit experiment, in which
electrons, say, impinge on a sheet with two slits
and are then allowed to fall on a screen. The wave function
is assumed to be a tightly peaked wave packet in the
direction perpendicular to the screen, but spreads
out in the direction parallel to it.
In the region
of the slits, we may assign a wave function
\begin{equation}
\psi  = \psi_1 + \psi_2
\end{equation}
to the electrons, where $\psi_1$ represents a wave emerging
from slit $1$, and $\psi_2$ represents a wave emerging
from slit $2$. The probability of hitting the screen at point
$x$ is then
\begin{eqnarray}
p(x)& = &| \psi (x) |^2
\cr
& = & | \psi_1  |^2 + | \psi_2 |^2  + \psi_1 \psi_2^* + \psi_1^* \psi_2
\end{eqnarray}
That is, the probability of hitting the screen at point $x$
is not the sum of the probabilities of the two separate
paths taken by the electron, as it would be in the classical
case. The probability sum rules are changed by the presence
of quantum interference.
Quantum effects are therefore characterized by a failure
of usual classical probability sum rules. Or differently put,
this corresponds to a failure of Boolean logic.

Note also the appearance of histories at this elementary
level. The interference effect arises as a result of the
fact that we are trying to talk about the properties of
the electron at both the screen {\em and} the slits, that is,
at two moments of time. This simple aspect is rarely mentioned
in elementary textbooks, but turns out to be significant,
as we shall see.

We see from this simple but important example that for a quantum
system to become classical, interference terms must be destroyed.
How does this come about? It is known in the double slit
experiment that if we measure the position of the electron close
to the slits, then the wave function will collapse into one of the
states $\psi_1$ or $\psi_2$ and the interference terms will go
away. We therefore expect that any physical mechanism that
constitutes some kind of physical measurement of the electron will
produce a similar effect. This is indeed the case. Suppose,
following the original calculation of Joos and Zeh we couple the
electrons to an environment \cite{JoZ}. This could be, for
example, a bath of photons. If the total wave function of the
system (electron) together with its environment is denoted by $ |
\psi_{S \E} \ra $, then the density operator of the system only is
\begin{equation}
\rho = {\rm Tr} | \psi_{S \E} \ra \la \psi_{S \E} |
\end{equation}
where ${\rm Tr}$ denotes a trace over the environment.
Such a density operator usually obeys a non-unitary
master equation. It typically has the form (in one dimension),
\begin{equation}
{ \partial \rho \over \partial t}
= { i \hbar \over 2 M} \left( { \partial^2 \rho \over \partial x^2}
- { \partial^2 \rho \over \partial y^2} \right)  - D (x-y)^2 \rho
\end{equation}
The important term is the last one on the right, which has the
effect of causing the density operator $\rho (x,y)$
to become approximately diagonal in position very rapidly.
The means that an initial state of the form
\begin{equation}
\rho = | \psi \ra \la \psi |
\end{equation}
where
\begin{equation}
| \psi \ra = | \psi_1 \ra + | \psi_2 \ra
\end{equation}
becomes essentially indistinguishable from the
classical mixture
\begin{equation}
\rho = | \psi_1 \ra \la \psi_1 | + | \psi_2 \ra \la \psi_2 |
\end{equation}
This is the phenomenon of decoherence, in its simplest form.

Consider now the question of how a quantum system, characterized
by its positions at a series of times, may become approximately
classical. Quantum theory supplies probabilities, so suppose
we use it to compute the probability
$ p (\a_1, t_1, \a_2, t_2, \cdots ) $ that the a particle is
in a spatial region $\a_1$ at time $t_1$, and then in a
spatial region $\a_2$ at times $t_2$, etc.  Then, ``approximately
classical'' means at least two things.
Firstly, it means that the probability is defined.
As we have seen, interference can prevent this from being
the case.
Secondly, it means that
the probability is strongly peaked when the regions $\a_1, \a_2,
\a_3 \cdots $ lie along a classical trajectory \cite{Har6}.

From this we see that histories provide an important mode of description
for discussing emergent classicality. It is therefore perhaps not
surprising that all of the above is most clearly formulated
in terms of the decoherent histories approach.

\section{The Decoherent Histories Approach}

The decoherent (or ``consistent'') histories approach was put
forward by Griffiths in 1984 \cite{Gri1} whose work was
substantially developed by Omn\`es \cite{Omn1,Omn2,Omn3}. It was
also discovered and developed, in part independently, by Gell-Mann
and Hartle \cite{GH1,GH2,Har1}. See Ref.\cite{TB} for a very extensive
bibliography on decoherent histories, and decoherence generally.

\subsection{The Formalism}

In the decoherent histories approach to quantum theory, a
quantum-mechanical history is characterized by an initial (pure or
mixed) state $\rho$, and by a time-ordered string of projection
operators,
\begin{equation}
C_{\au} =  P_{\a_n}(t_n)\cdots P_{\a_1}(t_1)
\label{1.1}
\end{equation}
Here, the projectors appearing in (\ref{1.1}) are in the Heisenberg
picture,
\begin{equation}
P_{\a_k}(t_k) = e^{\ih H(t_k-t_0)} P_{\a_k} e^{-\ih H(t_k-t_0)}
\label{1.2}
\end{equation}
and $\au$ denotes the string of alternatives $\a_1, \cdots \a_n$.
The projection operators $P_{\a}$ characterize the different
alternatives describing the histories at each moment of time. The
projectors satisfy
\begin{equation}
\sum_{\a} P_{\a} =1, \quad \quad P_{\a} P_{\beta} = \delta_{\a
\beta} \ P_{\a} \label{1.3}
\end{equation}
More generally, it is also of interest to work with histories in
which the so-called class operators (\ref{1.1}) are given by sums of
string of projections (which are not necessarily then equal to
strings of projections). Since the class operator (\ref{1.1}) is not in
fact a projection operator, it is reasonable to ask why this
operator should be used to describe a history. This  and many
other related issues have been considered in detail by Isham
and collaborators \cite{Ish}.

Probabilities are assigned to histories
of a closed system via the formula,
\begin{equation}
p(\a_1, \a_2, \cdots \a_n) = {\rm Tr} \left( C_{\au} \rho
C_{\au}^{\dag} \right) \label{1.4}
\end{equation}
Interference between pairs of histories is measured by the
decoherence functional
\begin{equation}
D(\au, \au') = {\rm Tr} \left( C_{\au} \rho C_{\au'}^{\dag}
\right) \label{1.5}
\end{equation}
Probabilities can be assigned to histories if and only if all
pairs of histories in the set obey the condition of {\it consistency},
which is that
\begin{equation}
{\rm Re} D(\au , \au' ) = 0 \label{1.6}
\end{equation}
for $\au \ne \au' $. This condition is equivalent to the requirement that
all probabilities satisfy the probability sum rules. These, loosely speaking,
are that the probability of history $\au$ or history $\bu$ (for two disjoint
histories), should be $p(\au) + p(\bu)$. Typically, realistic physical mechanisms
which bring about the consistency condition (\ref{1.6}) also cause the
imaginary part of $D(\au, \au')$ to vanish for $\au \ne \au'$ as
well, so it is of interest to work with the stronger condition of
{\it decoherence}, which is
\begin{equation}
D (\au , \au' ) = 0 \label{1.7}
\end{equation}
for $\au \ne \au'$.
The two conditions of consistency and decoherence have different consequences,
and we discuss them below.

When the decoherence condition is satisfied, it is straightforward to see
that
\begin{equation}
p (\au) = {\rm Tr} \left( C_{\au} \rho \right)
\label{1.7b}
\end{equation}
since we have $ \sum_{\au} C_{\au} = 1 $. Eq.(\ref{1.7b}) is clearly positive
when the decoherence condition is satisfied, but not generally not otherwise.
Goldstein and Page have turned this around and suggested that positivity
of the expression (\ref{1.7b}) may be used to select physically
viable histories \cite{GPa}. This is clearly weaker than the usual decoherence
condition. A variety of decoherence conditions related
to those discussed above were also discussed in Ref.\cite{Fin}.

\subsection{Approximate Decoherence}

Typically, the decoherence condition Eq.(\ref{1.7}) is satisfied
only approximately, raising the question of what approximate
decoherence actually means and how small the off-diagonal terms
really need to be. In this context, it is worth noting that the
decoherence functional satisfies the inequality,
\begin{equation}
\left| D (\au, \au') \right|^2 \le D (\au, \au) D( \au', \au')
\label{1.7b}
\end{equation}
as shown in Ref.\cite{DH}. A natural approximate decoherence
condition is then to insist that the decoherence functional
satisfies Eq.(\ref{1.7b}) but with a small number $\epsilon$ on
the right-hand side. It may be shown that this also guarantees
that most of the probability sum rules are satisfied to order
$\epsilon^{1/2}$ \cite{DH}. On the more general issue of
approximate decoherence, it has been suggested that an
approximately decoherent set of histories may be turned into an
exactly decoherent set by small distortions of the histories, for
example, by small distortions of the operators projected onto at
each moment of time \cite{DoK}. A closely related suggestion,
which has been worked out in some detail in particular cases, is
that approximately decoherent histories are those whose
predictions are well-approximated by a hidden variable theory,
such as the deterministic quantum theory of 't Hooft \cite{Hal}.
See also Ref.\cite{McE} for further considerations of approximate
decoherence.

\subsection{Path Integral Form}

It is also useful to note that for histories characterized by
projections onto position at different time, the decoherence
functional is very usefully expressed in terms of a path integral:
\begin{equation}
D (\au, \au') = \int_{\au} {\cal D} x \int_{\au'} {\cal D} y
\ \exp \left( \ih S[x(t)] - \ih S [y(t)] \right) \ \rho_0 (x_0, y_0)
\label{1.7b}
\end{equation}
The path integral is over a pair of paths $x(t), y(t)$ which are folded
into the initial state $\rho$ at $x_0, y_0$, wmeet at the final
time (and the final point is summed over), and at intermediate times
pass through the series of pairs of regions denoted by $\au$ and $\au'$.
This form is particularly useful for generalizations to situations which
involve time in a non-trivial way.

\subsection{Consistency}

Consistency, Eq.(\ref{1.6}), is an interesting condition, because, according to a theorem
of Omn\`es, systems described by a set of consistent histories from a
representation of classical logic \cite{Omn1}. That is, in a consistent set of
histories each history corresponds to a proposition about the properties
of a physical system and we can manipulate these propositions without
contradiction using ordinary classical logic. In simple terms, we
can {\em talk about} their properties. It is for this reason that the
decoherent histories approach may be thought of as supplying a foundation
for the application of ordinary reasoning to closed quantum systems.
In the Copenhagen interpretation, it was asserted that one could only
talk about {\em measured} quantities in an unambiguous way. In the decoherent
histories approach, the idea of a a measured quantity is replaced by the weaker
and more general idea of consistency.

An important example is the case of retrodiction of the past from
present data.
Suppose we have a consistent set of histories.
We would say that the alternative $\a_n$ (measured present data)
implies the alternatives
$\a_{n-1} \cdots \a_1 $ (unmeasured past events) if
\begin{equation}
p(\a_1, \cdots \a_{n-1} | \a_n ) \equiv
{ p(\a_1, \cdots \a_n )  \over p(\a_n) } = 1
\label{1.8}
\end{equation}
In this way, we can in quantum mechanics build a picture
of the history of the universe, given the present data and the
initial state, using only logic and the consistency of the
histories. We can meaningfully talk about the past properties of
the universe even though there was no measuring device there to
record them. This is one reason why the histories approach is of
interest in quantum cosmology.

\subsection{Decoherence and Records}

The stronger condition of decoherence is perhaps more interesting
since it is related to the existence of records -- some physical
mechanism existing at a fixed moment of time which is correlated
with the past history of the system. An example is a photographic
plate showing a particle track. In particular,
if the initial state is pure, there exist a
set of records at the final time $t_n $ which are perfectly
correlated with the alternatives $\a_1 \cdots \a_n $ at times $t_1
\cdots t_n $ \cite{GH2}. This follows because, with a pure initial
state $| \Psi \ra$, the decoherence condition implies that
the states $ C_{\au} | \Psi \ra $ are an orthogonal set.
It is therefore possible to introduce a projection operator
$ R_{\bu} $ (which is generally not unique) such that
\begin{equation}
R_{ \bu } C_{\au} | \Psi \ra = \delta_{\au \bu}
C_{\au} | \Psi \ra
\label{1.9}
\end{equation}
It follows that the extended histories characterized by the chain
$ R_{\bu} C_{\au} | \Psi \ra $ are decoherent,
and one can assign a probability
to the histories $\au$
and the records $\bu$, given by
\begin{equation}
p( \a_1, \a_2, \cdots \a_n ; \b_1, \b_2 \cdots \b_n )
= {\rm Tr} \left( R_{\b_1 \b_2 \cdots \b_n} C_{\au} \rho C_\au^{\dag} \right)
\label{1.10}
\end{equation}
This probability is then zero unless $\a_k = \b_k $ for all $k$, in
which case it is equal to the original probability  $ p(\a_1, \cdots
\a_n) $. Hence either the $\a$'s or the $\b$'s can be completely
summed out of Eq.(\ref{1.5}) without changing the probability,
so the probability for the histories can be entirely
replaced by the probability for the records at a fixed moment of
time at the end of the history:
\begin{equation}
p (\au ) = {\rm Tr} \left( R_{\au}  \rho (t_n ) \right)
= {\rm Tr} \left( C_{\au} \rho C_{\au}^{\dag} \right)
\label{1.11}
\end{equation}
Conversely, the existence of records $\b_1, \cdots \b_n $ at some
final time perfectly correlated with earlier alternatives $\a_1,
\cdots \a_n $ at $t_1, \cdots t_n $ implies
decoherence of the histories. This may be seen from
the relation
\begin{equation}
D(\au, \au' ) = \sum_{\b_1 \cdots \b_n }
{\rm Tr} \left( R_{\b_1 \cdots \b_n } C_\au \rho C_{\au'}^{\dag} \right)
\label{1.12}
\end{equation}
Since each $\b_k $ is perfectly correlated with a unique
alternative $ \a_k $ at time $t_k $, the summand on the
right-hand side is zero unless $\a_k = \a_k'$ (although
note that, as we shall see later, a perfect correlation
of this type is generally possible only for a pure initial
state).

There is, therefore, a very general connection between decoherence
and the existence of records. From this point of view, the
decoherent histories approach is very much concerned with
reconstructing possible past histories of the universe from records
at the present time, and then using these reconstructed pasts to
understand the correlations amongst the present records \cite{Har7}.

Some explicit models where the records may be explicitly
identified have been worked out. Ref.\cite{Hal5} showed how the
environment stores records of the particle's history in the
quantum Brownian motion model (in which the environment is a set
of harmonic oscillators). This was repeated for the case of
decoherence by a series of scattering processes in
Ref.\cite{HaDo}.

\subsection{The Non-Uniqueness of Retrodiction}

The above scheme contains an important subtlety that we now need
to discuss. This is that the retrodicted past is not in fact
unique: there are often many sets of consistent histories
associated with the same initial state and final measurement, and
what's more, can give conditional probability $1$ for different
complementary observables at an intermediate time. To see how this
happens, consider the following example (due to Omn\`es.) Suppose
we have a radioactive atom sitting at the origin which decays at
$t=0$, and therefore emits a particle in an outgoing spherical
wave state, and then hits a detector at time $t= t_2$. It is then
reasonable to ask if we can say anything about the system at an
intermediate time $t_1$.

We may analyze this using the decoherent histories approach.
The appropriate decoherence functional is,
\begin{equation}
D (\a_1, \a_2 | \a_1', \a_2 )
= {\rm Tr} \left( P_{\a_2} (t_2) P_{\a_1} (t_1) \rho P_{\a_1'} (t_1)
\right)
\label{1.13}
\end{equation}
Here, $P_{\a_2}$ is a projector onto the position of the detector,
$\rho = | \psi \ra \la \psi | $ is the spherical wave state,
and $P_{\a_1}$ is a projector
onto possible properties we may measure at the intermediate time
$t_1$. We consider two cases.

First, suppose that we project at
time $t_1$ onto a spatial region lying between the detector and
the origin. Then, one finds quite easily that the histories
are approximately decoherent, and secondly, that the conditional
probability $p(\a_1 | \a_2 ) $ of finding the particle at $\a_1$,
given that it was at $\a_2$ is approximately equal to $1$. That is,
if the detector clicks, then we may logically
deduce that the particle followed
a trajectory in the past, along the direct line from the origin
to the detector.

Second, suppose we project instead onto a completely different intermediate
state. The initial state is an outgoing spherical wave, and we there
expect that there will be consistent histories reflecting this fact.
Suppose we therefore project onto this possibility, using the projector,
\begin{equation}
P_{\a_1} (t_1) = | \psi (t_1) \ra \la \psi (t_1) |
\label{1.14}
\end{equation}
(together with its negation).
Then we again get consistency, and that $p(\a_1 | \a_2 ) = 1 $, from
which we would be inclined to say that the particle was in a spherical
wave state at time $t_1$. A third possibility would be to project
onto the momentum at $t_1$, which would lead to yet another set
of consistent histories, in which the value of momentum is predicted
with near certainty.

This illustrates that there are different sets of consistent histories
for the same physical situation, depending on which variables one would
like to talk about. Moreover, these different choices for $P_{\a_1} (t_1)$
do not commute, and this means that if we attempted to combine them in a
single set, the consistency condition would no longer be satisfied,
and we would not be allowed to make any logical deductions.
This feature of the formalism is essentially quantum-mechanical
complementarity, although it appears in a form which is for some
quite disconcerting. It raises the question as to whether one would
be able to assign definite values to non-commuting observables.
This, however, is excluded by a rule proposed by Griffiths \cite{Gri1}, which
states that all logical deductions about the system must be made
from the framework of a single consistent set.

The non-uniqueness of the retrodicted past then raises some questions
as to the value of the whole formalism, and what practical use it is.
It means that quantum theory, in the decoherent histories version of
it, does not uniquely tell us what ``actually happened'', although
to be sure, this feature of quantum theory is certainly known already
in various ways. See Refs.\cite{Omn3,DoK,GiR,Ana} for disucussions of these
issues.

The value of the formalism becomes clear when we ask what do we actually
do with the retrodicted past. The answer is that we use it to make other
predictions about the present. That is, we start from some present
data (measurements or cosmological observations), and, using consistency
or decoherence, we retrodict the past. We then use
the retrodicted past to make more predictions about the present.
The role of the histories, therefore, is that they are an intermediate
tool which helps us to identify correlations between data sets
at a fixed moment of time. For example, when we look at a photograph
of a particle track, we see a series of dots, existing at a fixed
moment of time, which appear to be correlated. The explanation of
their correlation is to be found in appealing to the past history
of the system that produced the dots. Differently put, by looking,
for example, at just three dots, we can use decoherence and retrodiction
to deduce that a particle passed through on an approximately
classical history. We then use the history to correctly deduce the
location of the remaining dots on the photographic plate.
In brief, therefore, histories are a useful tool to help us understand
present records \cite{Har7}.

\section{Quantum Brownian Motion Model}

We now briefly consider a particular model, namely the quantum
Brownian motion model. This model has been extensively studied in
the literature so only the briefest of accounts will be given here
\cite{DH}. The model consists of a particle of mass $M$ in a
potential $V(x)$ linearly coupled to an environment consisting of
a large bath of harmonic oscillators in a thermal state at
temperature $T$ \cite{CaL}. We consider histories of position
samplings of the distinguished system. The samplings are
continuous in time and Gaussian sampling functions are used
(corresponding to approximate projection operators). The
decoherence functional for the model is most conveniently given in
path-integral form:
\begin{eqnarray}
D[\bx(t), \by(t)] &=& \int {\cal D} x
{\cal D} y \ \delta (x_f -y_f) \ \rho(x_0, y_0) \cr & \times &
\exp \left( \ih S[x(t)] - \ih S[y(t)] + \ih W[x,y] \right) \cr  &
\times & \exp \left( - \int dt \ { (x(t) - \bx(t) )^2 \over 2
{\s}^2 } - \int dt \ { (y(t) - \by(t) )^2 \over 2 {\s}^2 } \right)
\label{4.1}
\end{eqnarray}
Here, $S$ is the action for a particle in a potential $V(x)$,
$\bx(t)$, $\by(t)$ are the sampled positions and $x_f$ and $x_0$
denote the final and initial values respectively. The effects of
the environment are summarized entirely by the Feynman-Vernon
influence functional phase, $W[x,y]$, given by,
\begin{eqnarray}
W[x(t),y(t)] & = & - \int_0^t ds \int_0^s ds' [ x(s) - y(s) ]
\ \eta (s-s') \ [ x(s') + y(s') ] \cr & + & i \int_0^t ds \int_0^s
ds' [ x(s) - y(s) ] \ \nu(s-s') \ [ x(s') - y(s') ]
\label{4.2}
\end{eqnarray}
The explicit forms of the non-local kernels $\eta$ and $\nu$
may be found, for example, in Ref.\cite{AnHa}. Here it is assumed, as is typical
in these models, that the initial density matrix of the total
system is simply a product of the initial system and environment
density matrices, and the initial environment density matrix is a
thermal state at temperature $T$. Considerable simplifications
occur in a purely ohmic environment in the Fokker-Planck limit (a
particular form of the high temperature limit), in which one has
\begin{eqnarray} \eta(s-s') &=  M\gamma \ \delta '(s-s')
\\ \label{4.3}
\nu(s-s') &= { 2 M \gamma k T \over \hbar } \ \delta (s-s')
\label{4.4}
\end{eqnarray}
where $\gamma$ is the dissipation. For convenience we will work in
this limit.
One can see almost immediately that the imaginary part of $W$,
together with the Gaussian samplings in Eq.(\ref{4.1}), will
have the effect of suppressing widely differing paths $\bx(t)$,
$\by(t)$. Indeed, the suppression factor will be of order
\begin{equation}
\exp \left( - { 2 M \gamma k T \s^2 \over \hbar^2 } \right)
\label{4.5}
\end{equation}
In cgs units $\hbar \sim 10^{-27}$ and $k \sim 10^{-16}$,
so $k T / \hbar^2 \sim 10^{40} $ if $T$ is room temperature.
Values of order $1$ for $M$, $\gamma$ and $\s$ therefore
lead to an astoundingly small suppression factor.
Decoherence through interaction with a thermal
environment is thus a very effective process indeed.

More precisely, one can approximately evaluate the functional
integral Eq.(\ref{4.1}). Let $X= (x+y)/2$, $\xi = x-y$, and use the smallness
of the suppression factor to expand about $\xi = 0$. Then the $\xi$
functional integral may be carried out with the result,
\begin{eqnarray}
D[\bx(t),\by(t)] & = & \int {\cal D} X \ W( M \dot X_0, X_0)
\ \exp \left( - \int dt \ { ( X- {\bx+\by \over 2} )^2 \over \s^2 }\right)
\cr & \times &
\exp \left( - \int dt \ { F[X]^2 \over 2 (\Delta F)^2 }
- i\hbar \int dt \ { (\bx -\by) F[X] \over 4 \s^2 (\Delta F)^2 }
\right)
\cr & \times &
\exp \left( - \int dt \ {(\bx -\by)^2 \over 2 \ell^2 } \right)
\label{4.6}
\end{eqnarray}
where
\begin{equation}
F[X] = M \ddot X + M \gamma \dot X + V'(X)
\label{4.7}
\end{equation}
are the classical field equations with dissipation, and
\begin{eqnarray}
(\Delta F)^2 &= { \hbar ^2 \over \s^2 } + 4 M \gamma k T
\label{4.8} \\
\ell^2 &= 2 \s^2 + { \hbar^2 \over 4 M \gamma k T }
\label{4.9}
\end{eqnarray}
$W(M \dot X_0, X_0)$ is the Wigner transform of the initial density
operator.

The decoherence width Eq.(\ref{4.9}) does not, in fact, immediately indicate
the expected suppression of interference, because the
temperature-dependent term will typically be utterly negligible
compared to the $\s^2$ term. The point, however, is that more precise
notions of decoherence need to be employed. One should check some of the
probability sum rules, or use the approximate decoherence condition
discussed in Ref.\cite{DH}, in which the sizes of the off and
on-diagonal terms are compared. This has not been carried out for the
general expression (\ref{4.6}), and in fact seems to be rather hard.
Satisfaction of the approximate
decoherence condition was checked for some special cases in Ref.\cite{DH}.
Still, one expects the standard to which
decoherence is attained to be of the order of the suppression factor
(\ref{4.5}), {\em i.e.}, very good indeed.

Now consider the diagonal elements of the decoherence function,
representing the probabilities for histories.
\begin{eqnarray}
p[\bx(t)] & =  & \int {\cal D} X \ W( M \dot X_0, X_0) \cr &
\times & \ \exp \left( - \int dt \ { ( X- \bx )^2 \over \s^2 } -
\int dt \ { F[X]^2 \over 2 (\Delta F)^2 }  \right) \label{4.10}
\end{eqnarray}
The distribution is peaked about configurations $\bx(t)$
satisfying the classical field equations with dissipation; thus
approximate classical predictability is exhibited. The width of
the peak is given by (\ref{4.8}). Loosely speaking, a given
classical history occurs with a weight given by the Wigner
function of its initial data. This cannot be strictly correct,
because the Wigner function is not positive in general, although
it is if coarse-grained over an $\hbar$-sized region of phase
space \cite{HalW}.

The width (\ref{4.8}) has clearly identifiable contributions from quantum
and thermal fluctuations. The thermal fluctuations dominate the
quantum ones when $8 M \gamma k T \s^2 \ >> \ \hbar^2$, which, from
(\ref{4.5}), is precisely the condition required for decoherence, as
previously noted.
Environmentally-induced fluctuations
are therefore inescapable if one is to have decoherence.
This means that there is a tension between the
demands of decoherence and classical predictability, both of which
are necessary (although generally not sufficient) for the emergence
of a quasiclassical domain \cite{GH2}. This tension is due to the fact that
the degree of decoherence (\ref{4.5}) improves with increasing environment
temperature, but predictability deteriorates, because the
fluctuations (\ref{4.8}) grow.
However,
the smallness of Boltzmann's constant ensures that  the fluctuations
(\ref{4.8}) will be small compared to $F[X]$ for a wide range of temperatures
if $M$ is sufficiently large.  Moreover, the efficiency of
decoherence as evidenced through (\ref{4.5}) is largely due to the
smallness of $\hbar$, and will hold for a wide range of
temperatures. So although there is some tension, there is a broad
compromise regime in which decoherence and classical predictability
can each hold extremely well.

One might wonder what sort of restrictions the uncertainty principle places
on the degree to which probabilities for histories may be peaked about a particular
history. For clearly it is not possible to be perfectly peaked because that would
mean definite values for positions at different times, which are non-commuting
operators. It turns out that the Shannon information,
\begin{equation}
I = \sum_{\au} p (\au ) \ln p (\au)
\end{equation}
provides a useful measure of the degree of peaking, and the uncertainty
principle appears as a lower bound on $I$, as shown in Ref.\cite{Hal2}.

\section{Quantum Classical Couplings}

An interesting further development of the previous section concerns
the construction of consistent theories describing the interaction
of classical and quantum systems. This section primarily follows Ref.\cite{Hal8a}.
(See also Ref.\cite{DiHa}.)
This is a question of interest
in a variety of different areas, in particular, in quantum field
theory in curved spacetime, where one is interesting in assessing
the effect a quantum field has on a classical gravitational field.
To be more precise, let us consider the simpler case of
a massive particle of mass $M$, position $X$ and velocity $\dot X$.
Suppose this particle, which we take to be well-described by
classical mechanics, comes into interaction with a
particle of mass $m$, which is sufficiently light that it
is predominantly quantum-mechanical and described by a state vector
$ | \psi \ra $. What happens to the classical particle under these
conditions?

With a simple linear coupling between these systems, the classical
particle obeys the equation of motion,
\begin{equation}
M \ddot X + V' (X) + \lambda x = 0
\label{5.1}
\end{equation}
and the state of the quantum particle obeys the equation
\begin{equation}
i \hbar { d | \psi \ra \over d t } = H_X | \psi \ra
\label{5.2}
\end{equation}
where $H_X$ denotes the quantum particle's Hamiltonian
in the presence of a classical external field $X$.
The problem with this system, however, is how to interpret
the quantity $x$ in Eq.(\ref{5.1}), which clearly should
be an operator because it describe a quantum system.
The simplest suggested resolution to this is to
insert a quantum-mechanical expectation value, and to
use instead of (\ref{5.1}) the equation
\begin{equation}
M \ddot X + V' (X) + \lambda \la \psi | x | \psi \ra = 0
\label{5.3}
\end{equation}
However, one would expect this prescription to yield
reasonable results only in a limited set of circumstances.
Indeed, it gives physically unreasonable results in
the interesting case when the quantum particle is in a
superposition of localized position states. Clearly in
that case the correct physical answer is that the classical
particle should ``see'' one or other of the localized
superposition states, and not some kind of averaged
position, which is what (\ref{5.3}) indicates.

An elementary extension of the results of the previous section
may be used to construct a more sensible coupled classical-quantum
system. One may start from the assumption that there are no
truly classical systems, only quantum systems that are approximately
classical under certain circumstances, as discussed in the previous
section. Consider therefore, a large particle of mass $M$ and
position $X$ coupled to an environment, to make it approximately
classical. Suppose also that the massive particle is again coupled
to the light particle, as in Eq.(\ref{5.1}). We now look for decoherent
histories of the massive particle.

It is easily shown that there are decoherent histories of positions of
the massive particle, whose probabilities are strongly peaked about
the equation of motion,
\begin{equation}
M \ddot X + M \gamma \dot X + V' (X) + \lambda \bar x =  0
\label{5.4}
\end{equation}
These are the classical equations of motion with dissipation,
as expected (and also with fluctuations, encoded in the width
of the peak). The interesting extra ingredient, however, is
the quantity $\bar x$ which is now not an operator, but a stochastic
c-number, for which the decoherent histories approach supplies
a probability distribution function $ p [\bar x (t) ] $.
This is a physically sensible result: the classical particle
responds in a stochastic way to the quantum system.
Furthermore, this construction gives the expected sensible
results for superposition states of the quantum particle.

\section{Decoherent Histories and Quantum State Diffusion}

For the particular, yet commonly realized situation in which
there is a natural split into system and environment,
it turns out that the decoherent histories approach is
closely related to the quantum state diffusion approach
to open quantum systems. We begin be briefly explaining
what this is.

Very many open quantum systems are accurately described by
the Lindblad master equation for the reduced density
operator $\rho$ \cite{Lin}. This is
\begin{equation}
{ d \rho \over dt} = - {i \over \hbar}  [H, \rho ]
- {1 \over 2} \sum_{j=1}^n \left(
\{ L_j^{\dag} L_j, \rho \} - 2 L_j \rho L_j^{\dag} \right)
\label{6.1}
\end{equation}
Here, $H$ is the Hamiltonian of the open system in the absence of
the environment (sometimes modified by terms depending on the $L_j$)
and the $n$ operators $L_j$ model the effects of the environment.
The Lindblad form is the most general possible evolution
equation preserving positivity, hermiticity and trace,
subject only to the physically useful assumption that the
evolution is Markovian.
For example, in the quantum Brownian motion model,
the master equation has a single
non-hermitian $L$ which is a linear combination of position and
momentum operators, and the Markovian approximation is valid
for reasonably high temperatures.

Whilst the master equation is the correct quantum-mechanical
description for many open systems, its solutions do not easily
yield the expected physical picture that may be directly compared
with experiments. Take again the case of quantum Brownian motion.
There, coarse observations of the system yield trajectories in
phase space following approximately classical paths with
dissipation and fluctuations. Yet an initial localized wavepacket
will spread indefinitely as a result of quantum and thermal
fluctuations under evolution according to Eq.(\ref{6.1}).
Differently put, the density operator really corresponds to an
ensemble of trajectories, yet an experiment measures an individual
trajectory of the system.

The quantum state diffusion picture, introduced by Gisin and
Percival \cite{GP}, aimed to expose the individual
trajectories contained in the density operator equation. In this
picture, the density operator $\rho$ satisfying (\ref{6.1}) is
regarded as a mean over a distribution of pure state density
operators,
\begin{equation}
\rho = M | \psi \ra \la \psi |
\label{6.2}
\end{equation}
where
$M$ denotes the mean (defined below), with the pure states
evolving according to the non-linear stochastic Langevin-Ito
equation,
\begin{eqnarray}
| d \psi \ra & = &-\ih H |\psi \ra dt  + {1 \over 2} \sum_j \left(
2 \la L_j^{\dag} \ra L_j - L_j^{\dag} L_j - \la L_j^{\dag} \ra
\la L_j \ra \right) | \psi \ra \ dt
\cr
& + & \sum_j \left( L_j - \la L_j \ra \right) | \psi \ra \ d \xi_j(t)
\label{6.3}
\end{eqnarray}
for the normalized state vector $| \psi \ra $.
Here, the $d \xi_j$ are independent complex differential random
variables representing a complex Wiener process. Their linear and
quadratic means are,
\begin{equation}
M [ d \xi_j d \xi_k^* ] = \delta_{jk} \ dt, \quad
M[ d \xi_j d \xi_k ] = 0, \quad M [ d\xi_j ] = 0
\label{6.4}
\end{equation}
This equation is very similar to the explicit modified versions
of quantum theory, such as that due to Ghirardi {\it et al}.\cite{GRW}, but here the motivation is
different, in that the stochastic equation is not proposed as a fundamental
equation.

An interesting feature of Eq.(\ref{6.3}) is that its solutions
tend to undergo some kind of localization in time, quite the opposite
of the master equation. For example, it may be shown that
in the quantum Brownian motion model, all initial states become
localized around a wave packet tightly peaked in phase space
after a very short time, and thereafter remain localized and
follow the classical equations of motion \cite{GP,HaZ}. Hence the quantum state
diffusion approach naturally provides the intuitive appealing
and experimentally correct picture of an individual trajectory.
Furthermore, one can also calculate a probability for each
trajectory.

It is therefore easily seen that both of the intuitive picture
and physically predictions of quantum state diffusion are in fact
essentially the same as those provided by the decoherent histories
approach. For the histories approach also naturally provides
the picture of an individual trajectory, and a probability for those
trajectories. This is argued in much greater detail in Ref.\cite{DGHP}.
Differently put, the solutions to quantum state diffusion
in some sense represent a single history from a decoherent set.
Also, QSD was put forward as a phenomenological picture of
open quantum systems. This connection with the decoherent histories
approach may be thought of as a more fundamental justification
of QSD phenomenology.

More general versions of this connection, involving other types of stochastic
unravelings have been discovered by Brun \cite{Bru1}.

\section{Hydrodynamic Equations}

Most models of decoherence, such as the quantum Brownian motion model
of Section 4, rely on an obvious separation of the whole closed
system in a distinguished subsystem, and the rest (the environment).
This is a reasonable assumption for a wide variety of physically
interesting situations. The decoherent histories approach, however,
does not rely on such a separation, and this is important, because
there are situations or regimes where such a split does not necessarily
exist. This then raises the question, in a large and possibly complex
quantum system, with no obvious system-environment split, what
are the variables that naturally become classical and what sorts
of classical equations of motion emerge from the underlying quantum
theory? Differently put, what is the most general possible derivation
of emergent classicality?

Gell-Mann and Hartle have argued that
one particular set of variables that are strong candidates for
the ``habitually decohering'' variables
are the integrals over small volumes of
locally conserved densities \cite{GH2}. These variables are distinguished by the existence
of conservation laws for total energy, momentum, charge, particle
number, {\it etc.} Associated with such conservation laws are local
conservation laws of the form
\begin{equation}
{\partial \rho \over \partial t} +{\bf  \nabla} \cdot {\bf j} = 0
\label{7.1}
\end{equation}
The candidate quasiclassical variables are then
\begin{equation}
Q_V = \int_{V} d^3 x \ \rho({\bf x})
\label{7.2}
\end{equation}
If the volume $V$ over which the local densities are smeared is
infinite, $Q_V$ will be an exactly conserved quantity. In quantum
mechanics it will commute with the Hamiltonian, and, as is easily
seen, histories of $Q_V$'s will then decohere exactly \cite{HLM}.
If the volume is finite but large compared to
the microscopic scale, $Q_V$ will be slowly varying compared to all
other dynamical variables. This is because the local conservation
law (\ref{7.1}) permits $Q_V$ to change only by redistribution, which
is limited by the rate at which the locally conserved quantity can
flow out of the volume. Because these quantities are slowly varying,
histories of them should therefore approximately decohere.
Furthermore, the fact that the $Q_V$'s are slowly varying may also be
used, at least classically, to derive an approximately closed set of
equations involving only those quantities singled out by the
conservation laws. These equations are, for example, the
Navier-Stokes equations, and the derivation of them is a standard
(although generally non-trivial) exercise in non-equilibrium
statistical mechanics \cite{For}.

One of the current goals of the decoherent
histories approach is to carry this programme through in detail.
A more detailed sketch of how this works was put forward in Refs. \cite{Hal6,Hal7}
and a variety of models and related aspects are described in Ref.\cite{hydro}.

\section{Spacetime Coarse Grainings}

Another class of questions to which the decoherent histories approach
adapts very well are those that involve time in a non-trivial way.
In particular, the arrival time and tunneling time problems have been
the subject of considerable recent interest, that interesting stemming
from the fact that quantum mechanics does not obviously supply
a unique and physically reasonable prescription for addressing
those problems. The decoherent histories approach offers yet another
way to analyzing these questions.

To focus ideas, consider the question, what is the probability
that a particle enters the spatial region $\Delta$ at any time,
during the time interval $[0, \tau]$? This has been address by
Yamada and Takagi \cite{YaT}, Hartle \cite{Har2} and Micanek and Hartle \cite{MiH},
most of these concentrating on the case where $\Delta$ is
the region $ x < 0 $.
The decoherence functional can be constructed using the path
integral form Eq.(\ref{1.7b}) as a starting point. The fact that
the question is non-trivial in time means that there is interference,
and the decoherence or consistency conditions are satisfied only
for very special choices of initial state, and the resultant
probabilities are rather trivial.
For example, in
the case where $\Delta $ is $ x<0$, the initial state must be
antisymmetric about $x=0$, and the probability of entering
the region is zero. In particular, this analysis does not
admit the interesting case of a wave packet starting in $x>0$
and approaching the origin. Moreover, it does not even give
results that have a sensible classical limit.

One can develop this approach further, and include an environment
to produce decoherence, along the lines of the quantum Brownian
motion model of Section 4 \cite{HaZa}. This then allows the assignment of
probabilities for a wide variety of initial states, with a sensible
and expected classical limit. However, they
are essentially the probabilities one might anticipate on the
basis of a classical Fokker-Planck equation for classical Brownian
motion, and moreover, the probabilities seem to depend on the
features of the environment.

At the present state of play, therefore, the results of the decoherent
histories analysis are not very striking. There is, however, space
for much more work to be done in this area. Furthermore, they appear
to be compatible with other approaches to the arrival time problem,
which involves, for example, time operators or detector models.
The fact that most initial states do not satisfy the consistency
condition (in the case of no environment) appears to be related
to the fact that paths in the path integral move in and out of
the region $\Delta $ many times, which in turn appears to be related
to the fact that the time operators proposed in other approaches
are not self-adjoint. Similarly, the dependence of the result on the
details of the environment seems to be related to the fact that,
in detector model approaches, the result depends on the form
of the detector. See Ref.\cite{Hal8} for a more detailed review of
these issues.

Finally, on the subject of temporal issues, it is worth noting
that the so-called temporal Bell inequalities \cite{PM}
may play in interesting role in the decoherent histories approach.
The temporal Bell inequalities are a set of inequalities on
certain probabilities or correlation functions which are derived
by assuming that there exists a valid probability distribution for
certain variables at a series of times. From the point of view of
the decoherent histories approach, this assumption is only true
when the histories are decoherent. The temporal Bell inequalities
may therefore provide an interesting way of characterizing decoherence
of histories, or the lack of decoherence, although this possibility
does not seem to have received much attention to date.

\section{Quantum Theory without Time}

An important extension of the ideas outlined in the previous
section is the further extension of the decoherent histories
approach to situations which do no involve specifying time
in any way whatsoever. For example, in classical mechanics
whether a particle follows a particular classical path.
Or we can ask whether a particle's path enters a given
region of space at stage along the entire trajectory.
Such questions are interesting and relevant for quantum
cosmology. There, the wave function of the system obeys
not a Schr\"odinger equation, but the Wheeler-DeWitt
equation, which has the form,
\begin{equation}
H \Psi = 0
\label{9.1}
\end{equation}
That is, the state function of the system is a zero energy
eigenstate. The form of this equation arises as a result
of reparametrization invariance, or more generally,
the four-dimensional diffeomorphism invariance of general
relativity. The problems of interpreting the solutions
to this equation are similar to the problems of interpreting
the Klein-Gordon equation as a first-quantized wave equation,
and have attracted a considerable amount of interest over
the years. Like the arrival time problem, there are a variety
of different approaches to it. The decoherent
histories approach is well-adapted to this problem.
This is partly because, as we shall see, the natural
reparametrization-invariant notion is an entire classical
trajectory, and the histories approach handles this notion
straightforwardly. We will briefly review the decoherent
histories analysis of the question. The detailed analysis of
the Klein-Gordon equation has been carried out in Ref.\cite{HaTh1}
and more general timeless models have been studied in Ref.\cite{HaTh2}.
We follow the latter quite closely.

To be precise, suppose we have an $n$ dimensional
configuration space with coordinates $\x = (x_1, x_2, \cdots x_n)$,
and suppose the wave function $ \psi (\x)$
of the system is in an eigenstate of the Hamiltonian,
as in Eq.(\ref{9.1}).
What is the probability of finding the system in a
region $\Delta$ of configuration space
{\it without reference to time}?
The classical case contains
almost all the key features of the problem and we concentrate on this.

We will consider a classical system described by a
$2n$-dimensional phase space, with coordinates
and momenta $ (\x, \p ) = (x_k, p_k) $, and Hamiltonian
\begin{equation}
H = {\p^2 \over 2 M} + V (\x )
\label{9.2}
\end{equation}
We assume that there is a classical phase space distribution
function $w (\p, \x ) $, which is normalized according to
\begin{equation}
\int d^n p \ d^n x \ w (\p, \x ) = 1
\label{9.3}
\end{equation}
and obeys the evolution equation
\begin{equation}
{ \partial w \over \partial t } = \sum_k
\left( - { p_k \over M} { \partial w \over \partial x_k}
+ { \partial V \over \partial x_k } { \partial w \over \partial p_k
} \right) = \{ H, w \}
\label{9.4}
\end{equation}
where $\{ \ , \ \}$ denotes the Poisson bracket. The interesting
case is that in which $w$ is the classical analogue of an energy
eigenstate, in which case $ \partial w / \partial t = 0 $, so the
evolution equation is simply
\begin{equation}
\{ H , w \} = 0
\label{9.5}
\end{equation}
It follows that
\begin{equation}
w( \p^{cl} (t), \x^{cl} (t) ) = w ( \p (0), \x (0) )
\label{9.6}
\end{equation}
where $\p^{cl}(t), \x^{cl}(t)$ are the classical solutions with
initial data $\p(0), \x(0)$, so $w$ is constant along the classical
orbits.

Given a set of classical solutions $ ( \p^{cl}(t), \x^{cl}(t) ) $,
and a phase space distribution function $w$, we are interested
in the probability that a classical solution will pass through
a region $\Delta $ of configuration space.
To see whether the classical trajectory $\x^{cl} (t)$
intersects this region, consider the phase space function
\begin{equation}
A( \x, \p_0, \x_0) =
\int_{-\infty}^{\infty} dt \ \delta^{(n)} ( \x - \x^{cl} (t) )
\label{9.7}
\end{equation}
(In the case of periodic classical orbits, the range of $t$ is
taken to be equal to the period). This function
is positive for points $\x $ on the classical trajectory
labeled by $\p_0, \x_0 $
and zero otherwise. It also has the property that
\begin{equation}
\{ H, A \} = 0
\end{equation}
so is a reparametrization-invariant observable. This is the mathematical
expression of the statement above that an entire classical trajectory
is a reparametrization-invariant notion.
Intersection of the classical
trajectory with the region $\Delta$ means,
\begin{equation}
\int_{\Delta} d^n x \ A (\x, \p_0, \x_0 ) > 0
\label{9.8}
\end{equation}
The quantity on the left is essentially the amount of parameter time the
trajectory spends in the region $\Delta $, so we are simply requiring
it to be positive.
We may now write down the probability for a classical
trajectory entering the region $\Delta $. It is,
\begin{equation}
p_{\Delta} = \int d^n p_0 d^n x_0 \ w (\p_0, \x_0 )
\ \theta \left( \int_{\Delta} d^n x \ A  - \e \right)
\label{9.9}
\end{equation}
In this construction, $\e$ is a small positive
number that is eventually sent to zero, and is included
to avoid possible ambiguities in the $\theta$-function
at zero argument. The $\theta$-function
ensures that the phase space integral is
over all initial data whose corresponding classical
trajectories spend a time greater than $\e$ in the
region $\Delta$.
Eq.(\ref{9.9}) is the desired formula for the probability
of entering the region $\Delta$ without regard to time.
All elements of it are reparametrization invariant.

At the meeting, a possible difficulty with Eq.(\ref{9.8}) was
pointed out by T.Brun. This is that for chaotic systems, the
trajectories visit every region in phase space, and it therefore
appears that every trajectory will intersect the region $\Delta$.
At present it therefore looks like the analysis presented here is
valid only for integrable systems. A more general version of this
analysis avoiding this difficulty is currently being sought, and
will be described elsewhere. A possible solution is that asking
whether the system intersects a given region of configuration
space is not in fact a useful observable for chaotic systems,
although there is no reason why there should not exist other types
of observables that are useful.

Turning now to the quantum theory, a decoherence functional
of the general form (\ref{1.5}) may be constructed.
Special attention is required for the inner product
(since the solutions to (\ref{9.1}) are typically not
normalizable in the usual sense), and also for the
class operators $C_{\au}$, which must be reparametrization
invariant and describe paths which pass through the region
$\Delta$. Also, an environment is typically required in
order to obtain decoherence.
These non-trivial aspects were discussed at length
in Ref.\cite{HaTh2}. The final result, in essence, is that in
the semiclassical limit, one obtains a formula
of the form (\ref{9.9}) in which the probability
function $w$ is replaced by the Wigner function of
the quantum state. Furthermore, this formula may be
rewritten in such a way that it coincides with earlier
heuristic analyses of the Wheeler-DeWitt equation (such
as the so-called WKB interpretation).

In summary, the decoherent histories approach successfully
adapts to genuinely timeless systems, such as the Wheeler-DeWitt
equation, although there is much more work to be done on
this question.

\section{Summary}

I have described the decoherent histories approach, its
properties and some recent applications and developments. These may be
summarized as follows:

\noindent $\bullet$ The decoherent histories approach is
essentially the Copenhagen approach to quantum mechanics,
but relies on a smaller set of assumptions. In essence,
it formalizes intuition.

\noindent $\bullet$ The approach gives a very comprehensive
account of emergent classicality in a variety of situations,
plus fluctuations about it, and couplings to quantum systems.

\noindent $\bullet$ It naturally accommodates situations where
there is no system-environment split.

\noindent $\bullet$ It readily generalizes to situations which
are non-trivial in time, or which do not involve time at all,
as we expect to be the case in quantum gravity.

\section{Acknowledgements}

I am very grateful to Thomas Elze to inviting me to take part in this most stimulating
meeting. I would also like to thank Todd Brun for useful comments on my talk
at the meeting.

\bibliography{apssamp}

\end{document}